\documentclass[a4paper,11pt]{article}
\usepackage[T2A]{fontenc}
\usepackage[cp1251]{inputenc}
\usepackage{amsmath,amsthm,amssymb}
\usepackage{graphics}
\usepackage{graphicx}
\begin{document}
\title{Non-equilibrium Spin-Hall Effect in Aluminum and Tungsten}
\author{Yu.N. Chiang, M.O. Dzyuba\\
\emph{B. I. Verkin Institute for Low Temperature Physics and Engineering,}\\
\emph{National Academy of Sciences of Ukraine.}\\
\emph{Nauky Ave. 47, Kharkov 61103, Ukraine}\\
\emph{E-mail:chiang@ilt.kharkov.ua}}
\date{}
\maketitle

\begin{abstract}
 The method proposed by us in [1], which eliminates obstacles in the application of electrical methods for studying the spin-Hall effect (SHE) by creating a spin unbalance, which generates a charge unbalance, using the form effect without using polarized injected current, is used in this work to study and compare SHE of different origin - internal (band) and external (structural and impurity). The internal SHE (ISHE) was studied on ultrapure single crystal samples of uncompensated (aluminum) and compensated (tungsten) metals. The influence of external factors on SHE was studied on polycrystalline and impurity aluminum samples. The investigations were carried out in the region of small magnetic fields, which ensure the symmetry of the conventional Hall effect (CHE), which makes it possible to study the chiral behavior of the spin Hall effect.
\end{abstract}
Earlier, Mott first formulated the conditions for the chiral asymmetry of the scattering of electron spins in the central force field of impurities in the presence of spin-orbit interaction.[2]. This served as the basis for the prediction of Dyakonov and Perel's  effect of curving trajectories of electrons with opposite spin polarization and their subsequent accumulation on opposite edges of the samples for nonmagnetic conductors [3], known as the spin Hall effect (SHE). In view of the lack of evidence of the effect for solids, interest in it was revived many years later only after the appearance of a number of experimental facts that pointed to the reality of its manifestation in solids [4 - 7]. This gave hope for the practical use of the spin-hall effect as an important tool for spintronics to manipulate spins, and the subsequent violent experimental and theoretical studies were aimed at the implementation and consideration of the Mott mechanism of spin polarization based on impurity (external) scattering [8]. Along with this, further theoretical studies of the effect indicated the possibility of an intrinsic (internal) spin polarization mechanism in the presence of spin-orbit interaction, which removes double spin degeneracy [9]. It was shown that under these conditions the external electric field can generate transverse additions of a different sign to the velocity of electrons with different signs of the quasimomentum, leading to spin currents of the opposite direction. In contrast to the Mott external mechanism, the internal mechanism of spin polarization, therefore, is only due to the band electronic structure of the metal and is not associated with impurity scattering [10, 11]. Using very pure metals with a low concentration of impurities, in this work we set as our goal to study the contribution of the internal mechanism of polarization of spins and spin currents and compare it with the scale of the influence of external factors on the magnitude of the spin Hall effect due to the spin-orbit interaction.

It should be noted that relativistic effects in their magnitude are effects of the following order of smallness, requiring a high degree of resolution of electrical signals, which we realized with the help of a high-precision measurement technique [12].

The main problem in the study of SHE has so far been the inability to use direct electric methods for measuring it because of the idea of symmetric and equal spin-currents of the opposite polarization transverse to the injection current $j_{x}$ caused by the spin-orbit interaction (see below). Therefore, in order to work under conditions of a nonequilibrium distribution of spins that make it possible to investigate SHE by electrical methods, one resorts to injecting knowingly spin-polarized carriers [7]. For this, as a rule, ferromagnetics are used whose ability to create a spin polarization of the current by the local magnetism of atoms is confirmed by observing in them an anomalous spin-Hall effect. The spin-polarized current obtained with the help of a ferromagnetic was then injected into the material under investigation, permitting one to study the SHE in it as an additive to the given nonequilibrium carrier state with respect to the charge and the spin of the system as a whole by means of electric methods [13, 14]. It is clear that the study of SHE by such indirect methods leaves a number of uncertainties in the interpretation of the results.

In order to avoid uncertainties in recognizing the contribution of SHE due to SO interaction (SOI), we proposed to create a nonequilibrium transverse distribution of spins and associated charges, without resorting to improper methods of current polarization. This can be done by making samples with an asymmetric shape of the cross-section, the characteristic size of which ($a\sim\sqrt{A}$, \emph{A} is cross-sectional area) is significantly larger than the mean free path of carriers $\ell_{c}$ and spin relaxation length $ l_{sf} $ [1].

\section{Formulation}
The physics of the appearance of a nonequilibrium spin regime with the given geometry of the sample cross section in a semiclassical language can be described as follows.
\begin{figure}[htb]
\begin{center}
  \includegraphics[width=15cm]{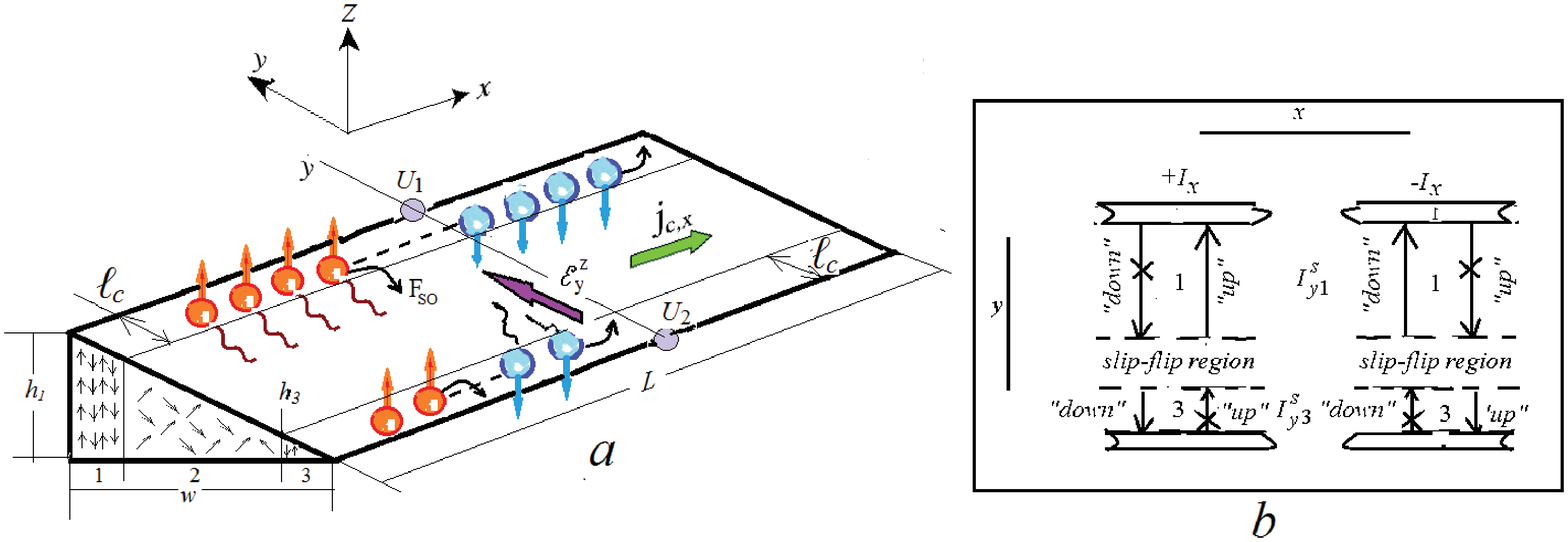}\\
  \caption{\emph{a})Schematic representation of a sample with an asymmetric cross section, leading to nonequilibrium dynamics and accumulation of spins in the presence of a spin-orbit interaction. 1 and 3 areas  refer to $ N_{1}$ and $ N_{3}$ numbers of carriers, respectively. The spin-flip region is symbolically marked by 2. \emph{b})Insert: The diagram of spin-currents in an asymmetric sample with a finite length of spin relaxation.}\label{1}
  \end{center}
\end{figure}

As is well known, the electrostatic potential difference in a metallic sample, including the transverse Hall voltage $U_{y}$ in the conventional Hall effect (CHE), is related to the geometric dimensions of the sample, indicating that $U_{y} $ is determined by the total rather than the specific the number of charges in the sample (in contrast to the Hall constant). In other words, for an asymmetric cross section of the sample, $U_{y}$ depends on the character of the distribution of all carriers along the cross section. If the corresponding gradient of the electrochemical potential $\mathcal {E}_{y}$ is established by the transverse current of all carriers under the action of the Lorentz force, as in CHE, then $U_{y}$ is determined simply by the mean section thickness $\bar{d}$ on axis \emph{z} passing through the center of gravity of the section ($U_{y}\sim\bar{d}^{-1}$). However, the dynamics of spins in a spin-orbit field, as noted, can not lead to the same nonzero result for $\mathcal{E}_{y}$, since it generates equal, oppositely directed spin streams. Indeed, if the spin-orbit interaction, removing the energy degeneracy in the spin [9], leads to a nonequilibrium dynamics of the spins in the momentum space due to the appearance of the additive to the SO field $\mathcal{E}_{z}^{s}$, induced by the $\delta\mathbf{v}\neq 0$ drift additive to the carrier velocity due to the current $j_{c,x}$ along the \emph{x} (Fig. 1\emph{a}), the deviation of electron spins for $ k_{y}> 0 $ up and $ k_{y} <0 $ down (\emph{k} - quasimomentum) will be antisymmetric and equal in magnitude, generating the equality of oppositely directed transverse currents relative to \emph{x} : $|+j_{y}^{s}|\equiv |-j_{y}^{s}|$.
\begin{figure}[htb]
\begin{center}
  \includegraphics[width=9cm]{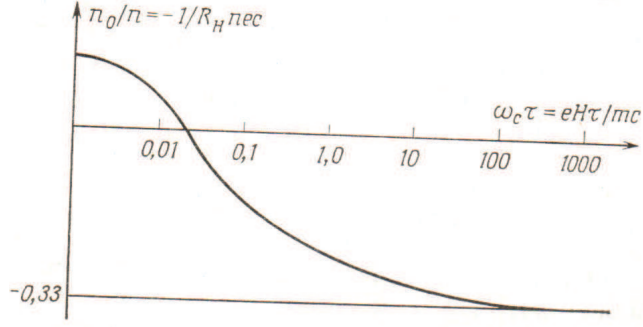}\\
  \caption{ The change in the sign of the carriers in Al in the region of small magnetic fields, caused by the redistribution of the contributions of the carriers with positive and negative effective masses [16,17].}\label{2}
  \end{center}
  \end{figure}

After establishing the equilibrium between the strength of the spin-orbital field $F_{so}$ and the gradient of the spin chemical potential $\nabla\mu_{y}^{s}$, arising between the spins of the opposite polarization ($F_{so}^{\pm} = e_{\mp} \nabla \mu_{y}^{s}$), the flow of spin-currents will result in the accumulation of charges with opposite spins on opposite edges of the sample in equal amounts, regardless of the cross sectional geometry, unless spin relaxation occurs in the sample. In this case, the appearance in the \emph{y} direction of the gradient of the spin chemical potential $\nabla\mu^{s}_{y}$ will not be accompanied by a charge imbalance and the gradient of the electrochemical potential in this direction remains zero. This means that the condition $\ell_{sf}\gg\sqrt {A_{yz}}$ makes it impossible to study SHE by electrical methods. However, if the dimensions of the sample in the cross section significantly exceed the mean free path ($\sqrt{A_{yz}}\gg\ell_{c}$), then it is reasonable to assume the inequalities $\ell_{c}\ll\ell_{sf}\ll\sqrt{A_{yz}}$, so there must be a "spin-flip" area (in Figure 1\emph{a} marked with the number 2), where the orientation of the moving spins stochastized, and the spin-currents of the oriented spins appear only where the spins are coherent, that is, in regions 1 and 3 in Fig. 1\emph{a} of thickness $\ell_{c}$ with an unequal number of carriers (spins) $N_{1}\sim A_{1}(\approx h_{1}\ell_{c})$ and $N_{3}\sim A_{3}(\approx \frac{1}{2}h_{3}\ell_{c}\approx \frac{1}{2} \frac{h_{1}}{w}{\ell_{c}}^{2})$. In this case, for an asymmetric cross section of the sample in the plane (\emph{yz}), as in Fig. 1\emph{a}, the total current of charges possessing spin dynamics in the presence of SOI is as follows:
\begin{equation}\label{1}
 \overline{I}_{y}^{s}=I_{y1}^{s}-I_{y3}^{s} = I_{y1}^{s}(1-\frac{I_{y3}^{s}}{I_{y1}^{s}}) = I_{y1}^{s}(1-\frac{N_{3}}{N_{1}}) = \sigma_{xy}^{s}E_{y}^{z}(1-\frac{A_{3}}{A_{1}}).
\end{equation}
\begin{figure}[htb]
\begin{center}
  \includegraphics[width=9cm]{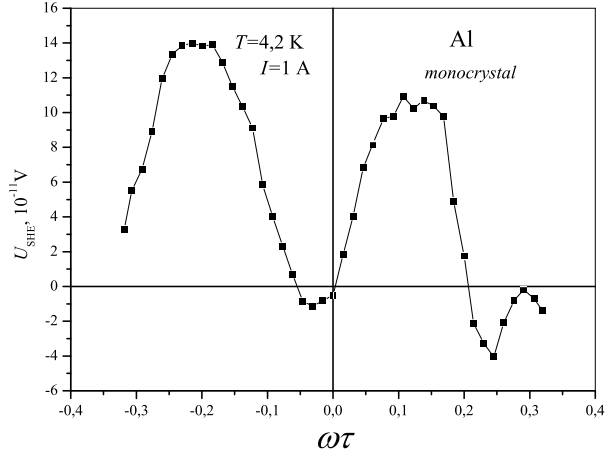}\\
  \caption{Nonequilibrium spin-Hall effect in macroscopic single-crystal aluminum sample with asymmetrical cross - sections in a perpendicular magnetic field.}\label{3}
  \end{center}
  \end{figure}
 It follows that $\overline{I}_{y}^{s}\neq 0$, if $A_{3}\neq A_{1}$, which is the condition for the subsequent nonequilibrium accumulation of spins at the sample boundaries in the direction  \emph{y} and establishing an appropriate charge potential difference $ U_{y}^{s} = U_{y1} - U_{y3}$ in this direction. It should be clear that the distinguishing feature of this transverse voltage will be the independence of its sign from the sign of $j_{c,x}$, since the direction of the charge imbalance does not depend on the polarization of the spins in the spin-currents, as seen from the circuit in Fig. 1\emph{b}. This allows us to distinguish and isolate $U_{y}^{s}$ in the presence of the CHE in the case of a magnetic field, since, unlike the conserved $ U_{y}^{s}$ sign, the  $U_{CHE}$ sign, given by the direction of the Lorentz force, commutes to the opposite one as each of the directions both the current $\mathbf{J}_{x}$ and $\mathbf{B}$ change, in particular $ \mathbf{B}_{z}$. Thus, the asymmetric shape of the sample makes it possible, generally speaking, to study SHE in its own nonequilibrium charge and spin states of the metal without resorting to preliminary polarization of the transport current by external methods, and the incompatibility of the behavior of $U_{\rm CHE}$ and $U_{\rm SHE }$ allows to confirm or deny the possibility of existence of such states when measuring the voltage $U_{y}$ in a magnetic field. Moreover, the effect of the measurable spin imbalance should manifest itself, apparently, in all those cases when the sample sizes cannot be maintained to within the electron mean free path in this material.
\begin{figure}[htb]
\begin{center}
  \includegraphics[width=9cm]{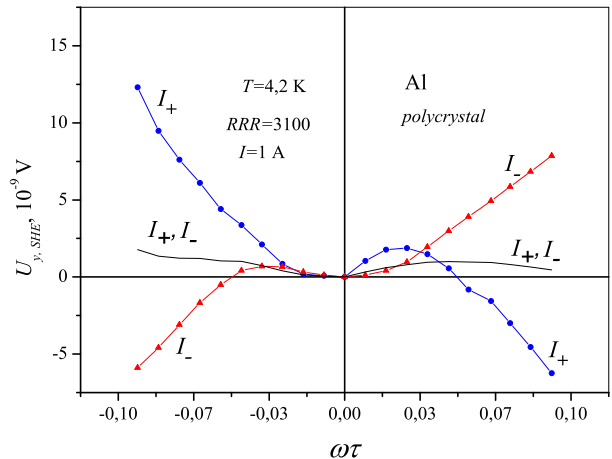}\\
  \caption{Nonequilibrium spin-Hall effect  in a macroscopic polycrystalline aluminum sample with asymmetrical cross - sections in a perpendicular magnetic field.~ $"I_{\pm}"$ are curves of the total transverse voltage along \emph{y}; $"I_{+},I_{-}"$ are curves of the nonequilibrium SHE.}\label{4}
  \end{center}
  \end{figure}

 The total voltage $U_{y}$ measured in a magnetic field $\mathbf{B}_{z}$ in the presence of current $\mathbf{J}_{x}$ on the Hall contacts $U_{y1}, U_{y2}$ will thus contain two components $U_{CHE}$ and $U_{SHE}$:
  \begin{equation}\label{1}
  U_{y}=U_{CHE} + U_{SHE} = \int_{0}^{w}\mathbf{E}_{CHE}\mathbf{j}_ydy + \int_{0}^{w}\mathcal{E}_{y}^{z}dy,
    \end{equation}
    where
    \begin{equation}\label{2}
      \mathbf{E}_{CHE} = (n^{\divideontimes}e)^{-1}[\mathbf{B}_{\pm z} \times \mathbf{J}_{\pm x}]
    \end{equation}
    and $\mathcal{E}_{y}^{z}$ is a Hall effective spin-dependent electric field [5, 15]
    \begin{equation}\label{3}
    \mathcal{E}_{y}^{z} = \frac{\alpha}{\ell_{c} k_{\rm F}}~[\mathbf{\hat{\sigma}}\times\mathbb{\nabla} \mu_{c}]_{y}.
    \end{equation}
  Here \emph{w} is the size of the sample along the axis \emph{y}; $ k_{\rm F} $ is the wave Fermi vector; $ \hat{\sigma}_{y} $ are the Pauli matrices; $ \alpha $ is the spin-orbit coupling constant, $ \mu_{c} $ is the chemical potential of the charge carriers, and $ n^{\divideontimes} $ is the effective concentration of charge carriers;~$\mathbf{j}_{y}$ is the unit vector of the y axis.
 \begin{figure}[htb]
\begin{center}
  \includegraphics[width=10cm]{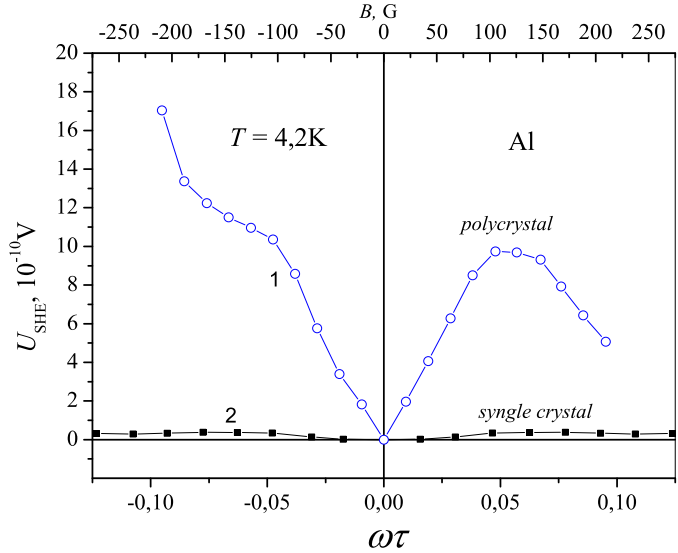}\\
  \caption{1 - Spin-Hall effect in a polycrystalline asymmetrical aluminum sample with \emph{RRR} = 3100 in a perpendicular magnetic field. 2 - SHE in a single-crystal asymmetrical sample of Al with $ RRR = 10^{4}$ for the same $I_{x}$ and orientation \emph{B}.}\label{5}
  \end{center}
\end{figure}
  \section{SHE in Aluminum}

  From the above it follows that the use of a magnetic field to identify the spin-hall effect, preferably for two reasons. First of all, due to the possibility of comparing the symmetric properties of SHE and CHE, which differ qualitatively, and in view of the very small magnitude of SHE as a relativistic second-order effect, depending on the magnitude of the magnetic field. Since, as noted, the chiral asymmetry of nonequilibrium SHE (NEQSHE) does not depend on the presence or absence of a magnetic field, and the symmetry of CHE depends, generally speaking, on the magnitude of the magnetic field, a reasonable compromise between the magnitude of CHE and the possibility of resolving SHE in normal metals can be achieved only in small effective magnetic fields  $\omega_{c}\tau\ll 1 ~(\omega_{c}~\mbox{is cyclotron frequency}, \tau~\mbox{is momentum relaxation time} $), where CHE in metals has almost complete symmetry about the directions and values of vectors $\mathbf{J}$ and $\mathbf{B} $.
  \begin{figure}[htb]
\begin{center}
  \includegraphics[width=10cm]{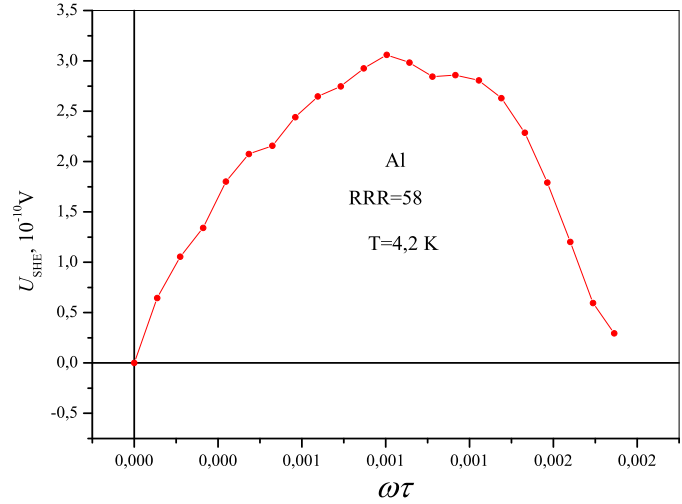}\\
  \caption{Spin-Hall effect in a polycrystalline asymmetrical dirty aluminum sample with \emph{RRR} = 58 in a perpendicular magnetic field.}\label{6}
  \end{center}
\end{figure}
In view of this, we worked in the range of effective magnetic fields  $\omega_{c} \tau \leq 0,1$, in which, moreover, the band characteristic of aluminum is manifested.

It is well known that it is  a change in the sign of the effective masses of carriers indicating the existence of alternative charge dynamics  and, as a consequence, the possibility of manifestation of the corresponding nonlinear behavior of NEQSHE in this field range. In Figure 2, this feature is shown as a change in the sign of the total carrier concentration with opposite carrier signs in the second ("hole") and third ("electronic") bonds, with $\tau /T_{c} = 0,01 (T_{c}=2\pi /\omega_{c})$, when during the $\tau$ relaxation time the carriers in both bonds have time make 0,01 full turn on oppositely directed trajectories of movement. Since a change in the carrier sign must change the sign of the charge voltage, the sign of the accumulated nonequilibrium charge voltage must also change, which in turn should result to a nonlinear feature in the behavior of the nonequilibrium SHE, playing a role, as one might expect, primarily as an identifier for the internal  SHE.

 To study this circumstance we have made the wedge-shaped aluminum specimens (see Fig. 1\emph{a}), that were cut  from a high-purity  single-crystal ingot (made by zone melting at the Volkhov Aluminum Plant named after SM Kirov) by an electric-spark method and subjected to chemical etching and polishing. The final resistivity of the single-crystal samples with was $\rho_{4,2 {\rm K}} \leq 1,6\cdot 10^{-10} {\rm Ohm\cdot cm}$ (RRR (Residual Resistance Ratio) $\geq 10^{4}$), so the magnetic fields of magnitude $0 < B < 300 \mbox{G}$ corresponded to the effective field domain $0 < \omega_{c}\tau < 0,1$. In [1], we reported on the detection in Al samples with such characteristics of the contribution to the transverse voltage $U_{y}$, which has all the features of NEQSHE: asymmetry with respect to the vectors $\mathbf{J}_{x}, \mathbf{B}_{\perp}$ and nonlinear behavior  in the $\omega_{c}\tau$ range of the above-mentioned band feature of this metal. From fig. 3, where this contribution is presented, it follows that the region of the manifested nonlinearity really coincides with the position of the corresponding feature in Fig. 2, confirming that we are dealing with the internal  SHE.

 Except internal SHE , we also studied NEQSHE in aluminum under the influence of some "external" factors, in particular, such as polycrystallinity and purity of the samples. A polycrystalline sample of aluminum of the same impurity composition as the single-crystal samples  was prepared by rolling a single crystal in such a way that the structure of the sample had a coarse-grained appearance with a grain size exceeding the mean free path ($\ell_{c}\approx 0, 03$ mm). Another sample was made of technical aluminum with a very small RRR = 58. (All data discussed in this paper are compared at the same value of the injected current 1A.)

  Fig. 4 shows the form of the total transverse voltage $U_{y}$ (curves $"I_{\pm}"$) and chiral component NEQSHE ($U_{\rm SHE}$) of this voltage (curves $"I_{+},I_{-}"$ in the range of values $\omega_{c}\tau$ corresponding to the same magnetic field interval of 0–300 G as in the measurement of single-crystal samples. In fig. 5 the $U_{\rm SHE}$ contribution is shown together with the similar contribution for a single-crystal sample shown in fig. 3. It can be seen that in comparison with the latter, the spin-hall effect in polycristalline sample  with the same current value increased by an order of magnitude, also exhibiting nonlinear behavior precisely in that region of small fields ($\omega_{c}\tau < 1$), where the sign of the effective mass of the carriers change  as in Fig. 2. In view of arbitrary grain shapes it can be expected  that the increase in the SHE in comparison with the case of the two boundaries is associated, in particular, with the sequential addition of the voltages spin-accumulated at the grain boundaries. As for the dirty sample, for which $U_{\rm SHE}$ is shown in fig. 6, then here we also observe an increased spin-Hall effect and its nonlinear behavior is in the same interval $\omega_{c}\tau < 1$.
 \begin{figure}[htb]
\begin{center}
  \includegraphics[width=5cm]{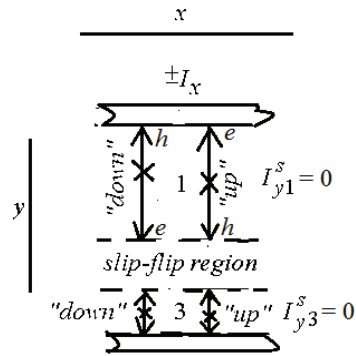}\\
  \caption{The spin-current diagram in a sample of an asymmetrical shape of a compensated metal with a finite spin relaxation length.}\label{7}
  \end{center}
\end{figure}

  Obviously, the spin-orbit coupling coefficient ($\alpha$ in expression (4)) can be correctly estimated only from measurements of the internal SHE, and in accordance with the values of the effect in Fig. 3, in order of magnitude, as noted in [1], it is within $\approx 1.13\cdot 10^{-5}$, which is 2 orders of magnitude smaller than the earlier estimates (see [8, 13]). It can be assumed that external factors, increasing the effect, as in Figure 4, cannot give an objective idea of the value of $\alpha$.
   \section{SHE in Tungsten}
The NEQSHE scenario considered above for the asymmetric geometry of a sample is applicable only to a metal with a predominant contribution of one type of carrier, which is, in particular, Aluminum. However, one can expect that in a two-band metal with equal electron and hole concentrations, a nonequilibrium spin state will not arise for any geometry of the samples. One of such metals is, as is known, tungsten. In a zero magnetic field and at very small values of it ($ 0 <\omega_{c}\tau\ll 1 $), the spin-currents of holes and electrons in the compensated metal should be equal and antisymmetric (Fig. 7), preventing the occurrence of NEQSHE.

 In Fig. 8 shows the spin-Hall voltage in an asymmetric sample of single-crystal tungsten with $RRR = 10^{4}$, isolated from the measured total transverse voltage $U_{y}$ (in the inset) according to the expression (2) in that magnetic field interval ($\omega\tau <1$), where the symmetry of $U_{\rm CHE}$ is valid. For comparison, the curve for a single-crystal sample of aluminum from fig. 3 is also shown. Can be seen that in tungsten there is no SHE in a comparable range of magnetic fields, at least in the scale of instrumental voltage resolution. A noticeable dependence on the magnetic field begins to appear at $\omega\tau\sim 1$, that is, when a symmetry breaking of CHE is possible due to the manifestation of anisotropy of the Fermi surface of tungsten.
\begin{figure}[htb]
\begin{center}
  \includegraphics[width=10cm]{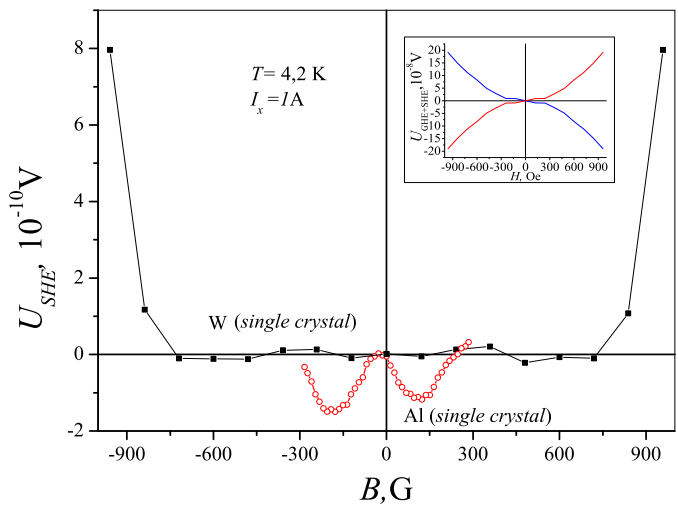}\\
  \caption{SHE in a tungsten sample with a wedge-shaped cross-section and a finite length of spin relaxation. The SHE curve from Fig. 3 for the sample Al  of similar shape is also shown for comparison there. The inset shows the initial total voltage measured in the transverse direction to the injected current in small perpendicular magnetic fields in a tungsten sample.}\label{8}
  \end{center}
\end{figure}
\section{Conclusion}
 Direct electrical measurements of the relativistic spin Hall effect due to spin-orbit interaction were carried out for the first time in normal aluminum and tungsten metals using unpolarized injection current. In particular, samples of these metals with a carrier mean free path of not less than $ 10 ^ {3} $ times greater than that used in SHE studies were studied. Since measurements of SHE by electrical methods are possible only under conditions of spin unbalance, the latter was created by the asymmetric shape of samples with dimensions that were known to exceed the length of the spin relaxation. The accompanying charge imbalance allowed for the first time to observe the chiral asymmetry of the internal spin Hall effect  in aluminum and tungsten. In particular, the effective magnetic field interval $0 <\omega\tau <0.1$ is used to identify ISHE, where the effective masses of dynamic carriers, and with them the effective  concentration, in Al change sign, and CHE in normal metals has almost complete symmetry. The nonlinear behavior of the transverse chiral voltage in Al found in this field region can be considered the first direct observation of the internal (band) SHE, confirming the existence of nonequilibrium spin dynamics in momentum space. The absence of SHE in compensated nonmagnetic tungsten metal also indirectly indicates this. The study of the influence of external factors on SHE, such as structural defects (polycrystallinity) and impurities, revealed a significant increase in the effect compared with the value of the internal SHE, which, at least in the case of impurities, can be attributed to the influence of the Mott mechanism of spin polarization. The value of the spin-orbit coupling coefficient obtained by us in aluminum is two orders of magnitude smaller than that estimated from indirect measurements of SHE in disordered metallic nanostructures [11].

We express our gratitude to Dr. Sologub S.V. (Institute of Physics NAS of Ukraine) for providing samples of ultrapure tungsten.

 \end{document}